\documentclass{emulateapj}
\usepackage{epsf,graphics}
\usepackage{subfigure,graphicx}
\usepackage{bm}
\usepackage{mathrsfs}
\usepackage{epstopdf}

\begin{document}

\title{Robust velocity dispersion and binary population modeling of the ultra-faint dwarf galaxy Reticulum II}

\author{Quinn E. Minor}
\affil{Department of Science, Borough of Manhattan Community College, City 
University of New York, New York, NY 10007, USA}
\affil{Department of Astrophysics, American Museum of Natural History, New York, NY 10024, USA}
\email{qminor@bmcc.cuny.edu}
\author{Andrew B. Pace, Jennifer L. Marshall, Louis E. Strigari}
\affil{Department of Physics and Astronomy, Mitchell Institute for Fundamental Physics and Astronomy, Texas A\&M University, College Station, TX 77843, USA}


\begin{abstract}
We apply a Bayesian method to model multi-epoch radial velocity 
measurements in the ultra-faint dwarf galaxy Reticulum II, fully accounting for the effects of binary orbital motion and systematic offsets between different 
spectroscopic datasets. We find that the binary fraction of Ret II is higher 
than 0.5 at the 90\% confidence level, if the mean orbital period is assumed to 
be 30 years or longer. Despite this high binary fraction, we infer a best-fit
intrinsic dispersion of 2.8$_{-1.2}^{+0.7}$ km/s, which is smaller than previous estimates, but still indicates Ret II is a dark-matter dominated galaxy. We likewise infer a $\lesssim$ 1\% probability that Ret II's dispersion is due to binaries rather than dark matter, corresponding to the regime $M_{\odot/L_\odot} \lesssim$ 2. Our inference of a high close binary fraction in Ret II echoes previous results for the Segue 1 ultra-faint dwarf and is 
consistent with studies of Milky Way halo stars that indicate a high close 
binary fraction tends to exist in metal-poor environments.

\end{abstract}


\section{Introduction}\label{sec:intro}

Ultra-faint dwarf (UFD) galaxies comprise the extreme end of the galaxy 
luminosity function, being the faintest and most dark matter dominated galaxies 
known.  Their high mass-to-light ratios \citep{simon2007}, paucity of gas and extremely low 
metallicities \citep{kirby2008} indicate they host very old and sparse star populations, with the 
bulk of their star formation occurring before the epoch of reionization and 
being quenched not long after \citep{brown2012}. As a result of their high dark matter content, 
UFD's are excellent laboratories for understanding the nature of dark matter in 
several respects: their expected abundance and density profile depends 
sensitively on the particle nature of dark matter, for example if it is warm \citep{lovell2014} or has a significant cross-section for self-interactions \citep{elbert2018,rocha2013}. In addition they are 
excellent targets to search for the products of dark matter self-annihilation, 
as a result of their low astrophysical backgrounds \citep{ackermann2015}. The study of UFD's has been reinvigorated in the past few years by the recent discovery of several new 
candidate dwarf galaxies by the Dark Energy Survey \citep{Bechtol2015ApJ...807...50B, Koposov2015ApJ...805..130K, Drlica-Wagner2015ApJ...809L...4D, Kim2015ApJ...808L..39K, Luque2016MNRAS.458..603L, Luque2017MNRAS.468...97L, Luque2017arXiv170905689L}.

The estimated high dark matter content of ultra-faint dwarf galaxies hinges on 
an accurate measurement of the line-of-sight velocity dispersion of their 
constituent stars, with dispersions typically in the range of 1-6 km/s \citep{simon2007}.  
However, their low intrinsic dispersions make the task difficult for a few 
reasons. First, in many cases they are too distant for radial velocities of 
main sequence stars to be measured by present instruments, leaving only the red 
giants in the sample.   Second, spectrographs have systematic 
floors which limit the measurement of stellar velocities to greater than $\sim 0.2-2.2$ km/s.
Finally, the dispersions are low enough that the orbital motion of close 
binary systems can inflate them significantly \citep{spencer2017,McConnachie2010ApJ...722L.209M,wilkinson2002}.

Of these systematics, binaries are perhaps the most worrisome, because unexpectedly high radial velocity variations have been detected in several UFD's. This occurred, for example, among RGB stars in the Segue 1 sample, which did not inflate its dispersion dramatically due to a large sample of main sequence turnoff stars \citep{Simon2011ApJ...733...46S}. An initial velocity dispersion measurement of Triangulum~II \citep{Kirby2015ApJ...814L...7K, Martin2016ApJ...818...40M} had 1-2 unresolved binaries that inflated its dispersion \citep{Kirby2017ApJ...838...83K}. Similarly, the dispersion of Bo\"{o}tes~II \citep{Koch2009ApJ...690..453K} is inflated due to a known binary  \citep{Ji2016ApJ...817...41J}, and Carina~II's dispersion would likely have been inflated if multi-epoch spectroscopy had not been obtained \citep[see one epoch row of Table 3 in ][]{Li2018ApJ...857..145L}. In classical dwarf spheroidals, measured binary fractions appear to vary considerably, with some being higher and others lower than the exected value for Milky Way field binaries \citep{Minor2013ApJ...779..116M,spencer2017,olszewski1996}. Besides dwarf galaxies, within the Milky Way there is mounting evidence that the fraction of close binaries is indeed higher in low metallicity systems \citep{Moe2018arXiv180802116M,badenes2018}. Thus, there is a very real possibility that the measured dispersions of many ultra-faint dwarfs are significantly inflated due to binary motion; indeed, some may turn out to be diffuse globular clusters in disguise.
    
The most direct way to correct the dispersions of UFD's for binary motion is to 
perform spectroscopic follow-up over two or more epochs, and model the binary 
contribution to the radial velocities directly. This strategy was carried out 
in the case of Segue 1 \citep{Simon2011ApJ...733...46S,Martinez2011ApJ...738...55M}. While spectroscopic follow-up has been recently performed 
for a few of the recent DES dwarfs, including Reticulum II, the binary modeling 
is complicated by the fact that radial velocity measurements have been made 
using different instruments and different methods of analysis \citep{simon2015,koposov2015,Ji2016ApJ...817...41J,roederer2016}. This gives rise to systematic velocity offsets between the different datasets which can masquerade as binary variability. As a result, any attempt to model the binary component of 
velocity variations should include systematic velocity offsets between datasets
as model parameters.

In this paper we constrain the velocity dispersion of the ultra-faint dwarf 
Reticulum II, including the effects of binary orbital motion and systematic 
offsets between datasets, in a Bayesian analysis. By constructing 
the binary likelihood for each star and comparing it to the single-star 
likelihood, a probability of binarity can be inferred that does not rely on 
definitively identifying an individual star as a binary. The approach is 
similar to that employed for Segue 1, except that in addition we include 
systematic offset parameters to be inferred alongside the binary population 
parameters; hence, any degeneracy between the intrinsic dispersion, 
instrumental systematics, and binarity can be inferred in a consistent manner.

\begin{figure*}[ht]
	\centering
	\subfigure[S15 vs. K15]
	{
		\includegraphics[height=0.4\hsize,width=0.48\hsize]{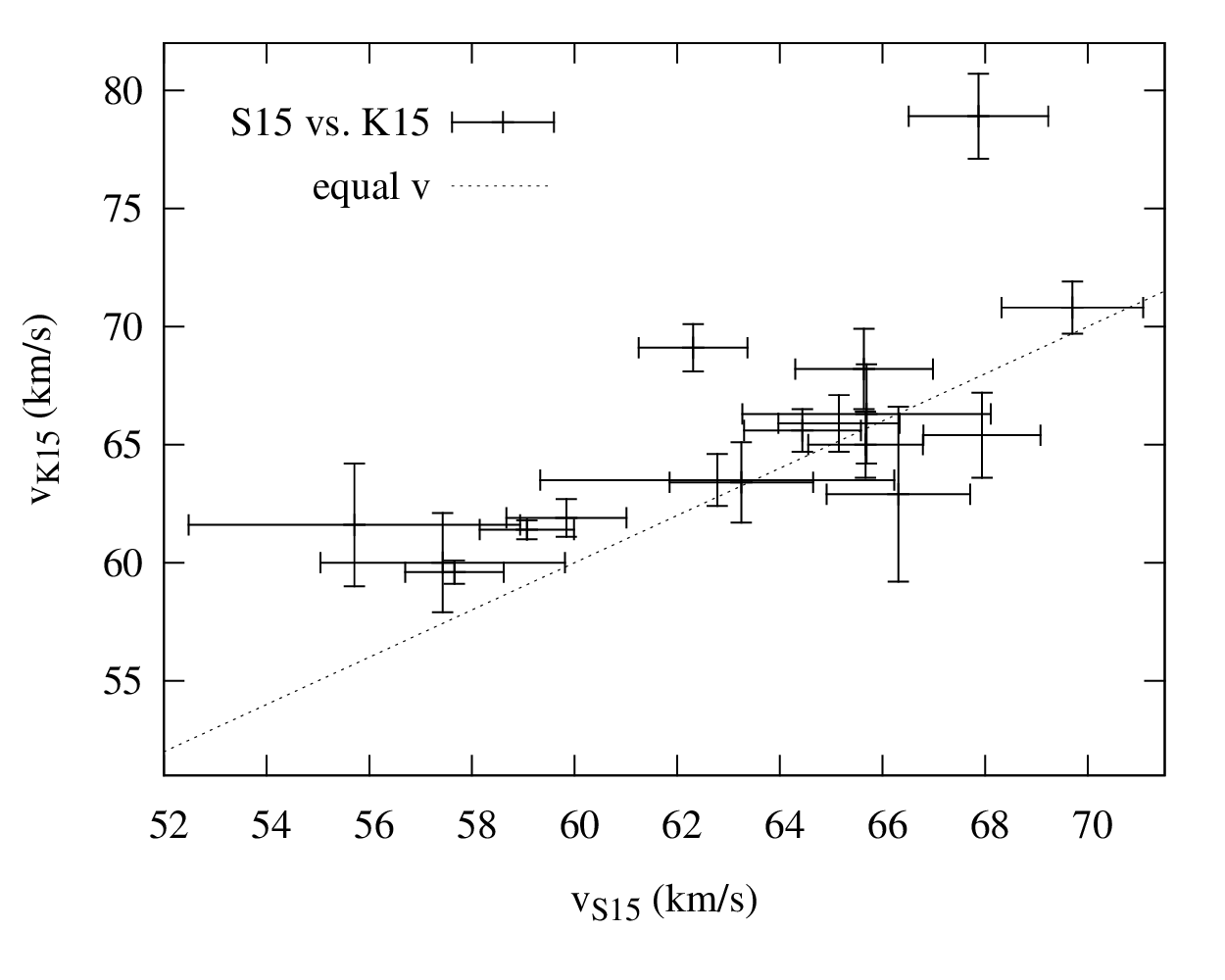}
		\label{fig:s15_vs_k15}
	}
	\subfigure[J16 vs. S15, K15]
	{
		\includegraphics[height=0.4\hsize,width=0.48\hsize]{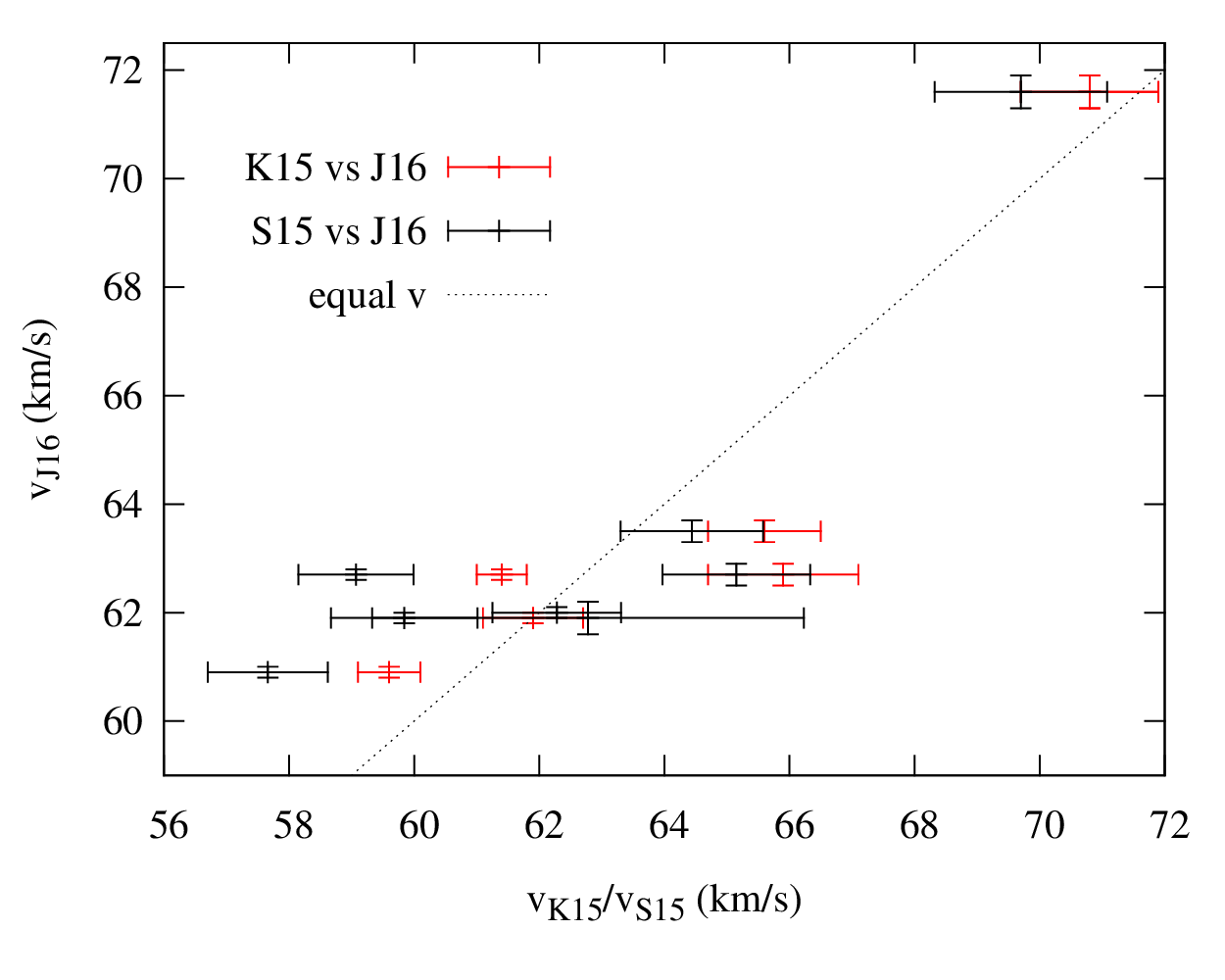}
		\label{fig:j16_vs_others}
	}
	\caption{Comparison of velocity measurements between datasets for stars that 
are common to both datasets. The error bars represent 1-sigma errors for each 
dataset.  The dashed line corresponds to equal velocities being measured, e.g.  
in (a) it represents $v_{S15} = v_{K15}$. Note that the S15 and K15 
measurements are separated by only 18 days, and therefore binary orbital motion 
is unlikely to account for the majority of offsets between the two datasets in 
figure (a).}
\end{figure*}

\section{Bayesian Method: correcting for binaries}\label{sec:method}

In order to correct the velocity dispersion of Reticulum II for binary orbital 
motion, we follow the method used in \cite{Martinez2011ApJ...738...55M} and \cite{Minor2013ApJ...779..116M}: we construct a 
multi-epoch likelihood for binary stars and use this to infer posterior probabilities in the galaxy kinematic parameters as well as parameters describing the binary population. Since this approach has been described 
in previous papers, we will give a brief summary here and describe the 
differences in our approach compared to previous papers.

Suppose a star of absolute magnitude $M$ has a set of $n$ velocity measurements 
$\{v_i\}=\{v_1,\dots,v_n\}$ and errors $\{e_i\}$ taken at the corresponding 
dates $\{t_i\}$. For readability, when writing probability distributions we 
will suppress the brackets that denote sets of measurements (e.g., $P(\{v_i\}) 
\rightarrow P(v_i)$). Since the velocities at different epochs in the Ret II 
dataset were measured using different spectrographs, we will allow for each 
epoch to have a possible systematic velocity offset $\lambda_i$. 
Thus, our likelihood will be written in terms of the corrected velocities 

\begin{equation}
v_i' = v_i - \lambda_i.
\end{equation}
    
For some epochs we will fix the corresponding systematic offsets to zero, while 
for others we will vary the offsets as free parameters. The set of systematic 
offsets to be varied will be denoted as $\mathcal{S} = 
\{\lambda_1,\lambda_2,\ldots\}$.  Our velocity likelihood then becomes,
\begin{eqnarray}
\lefteqn{\mathcal{L}(v_i|e_i,t_i,M;\sigma,\mu,B,\mathcal{S})} \nonumber \\
& ~ = & ~ (1-B)\mathcal{L}(\Delta v_{ij}',e_i)\frac{e^{-\frac{(\langle v'\rangle - \mu)^2}{2(\sigma^2 + e_m^2)}}}{\sqrt{2\pi(\sigma^2 + e_m^2)}} ~ ~ ~ ~ ~ ~ ~ ~ ~ ~ ~ ~ ~ ~ ~ ~ \nonumber \\
& & ~ ~ ~ ~ ~ ~ ~ ~ ~ ~ ~ ~ ~ ~ ~ ~  + ~  B\mathcal{L}_b(v_i|e_i,t_i,M;\sigma,\mu,\mathcal{S}),
\label{eq:full_binary_likelihood}
\end{eqnarray}
where the first term gives the likelihood for non-binary stars, and the second 
is the binary star likelihood. Note that the non-binary likelihood has been factorized into a likelihood in the star's mean velocity $\langle v'\rangle$, and a likelihood in terms of velocity $\emph{differences}$, denoted by $\mathcal{L}(\Delta v_{ij}',e_i)$; the latter is given by
\begin{eqnarray}
\lefteqn{
\mathcal{L}(\Delta v_{ij}',e_i) ~ ~ = ~ ~ \frac{\sqrt{2\pi e_m^2}}{\prod_{i=1}^n \sqrt{2\pi e_i^2}} } \nonumber\\
& \times & \exp\left\{-\frac{1}{4}\sum_{i,j=1}^n\frac{(v_i'-v_j')^2}{e_i^2 + e_j^2 + e_i^2 e_j^2\left(\sum_{k\neq i,j}\frac{1}{e_k^2}\right) } \right\}. \\
\nonumber
\label{eq:n_factor}
\end{eqnarray}

The last term in the denominator of the exponent is implicitly zero when $n = 
2$. To generate the likelihood for binary stars (the second term in equation 
\ref{eq:full_binary_likelihood}), we first generate the binary likelihood in 
the center-of-mass frame of the binary system, denoted by $P_b(v_i' - v_{cm}|e_i,t_i,M)$.  This is done by running a Monte Carlo simulation where the binary properties are drawn from distributions similar to those observed in the solar neighborhood (see \citealt{Minor2010ApJ...721.1142M} for details on the orbital parameters and their assumed distributions), and binning the resulting stellar velocities over a table of $v_{cm}$, $\lambda_i$ values. To ensure accuracy of the resulting likelihood, points are added to the bins until the binary likelihood (or rather, its integral over $v_{cm}$) converges to within a specified tolerance. For each likelihood evaluation, given values of the galaxy's systemic velocity $\mu$ and dispersion $\sigma$, we interpolate in the offset parameters $\lambda_i$ and integrate over the tabulated $v_{cm}$ values as follows:

\begin{equation}
\mathcal{L}_b(\sigma,\mu,\mathcal{S}) = \int_{-\infty}^{\infty} P_b(v_i'-v_{\mathrm{cm}}|e_i,t_i,M) \frac{e^{-\frac{(v_{\mathrm{cm}} - \mu)^2}{2\sigma^2}}}{\sqrt{2\pi\sigma^2}} dv_{\mathrm{cm}}.
\label{eq:i_integral}
\end{equation}

From a hierarchical modeling point of view, this can be thought of as 
marginalizing over $v_{cm}$ for each star, assuming a Gaussian prior whose 
hyperparameters $\mu$, $\sigma$ will be determined at the next level of 
inference (indeed, this has already been done implicitly for the nonbinary 
likelihood to give the Gaussian factor in eq.~\ref{eq:full_binary_likelihood}).

As in \cite{Martinez2011ApJ...738...55M}, we assume a log-normal period distribution for the binaries; however, we do not vary the period distribution parameters of 
the underlying binary population, since the constraints would be very poor given 
the sample size. However, for the sake of comparison we will test two models: 
one in which the period distribution is similar to Milky Way field binaries 
with a mean period of $\sim$180 years \citep{duquennoy1991}, and one in which the mean period is $10^4$ days $\approx$ 27 years. As we discuss in Section \ref{sec:results}, the latter choice is motivated by the mean period inferred in low-metallicity halo stars in \cite{Moe2018arXiv180802116M}. In both cases, the width of the log-period distribution is assumed to be $\sigma_{\log P}=2.3$, similar to that of binaries in the solar neighborhood as determined by \cite{duquennoy1991} and independently by \cite{raghavan2010}.

\section{Data}

We make use of three different radial velocity datasets from Ret II, composed 
entirely of red giant stars with high probability of membership.  The bulk of 
the velocity measurements come from \cite{simon2015} using the Magellan/M2FS 
spectrograph (hereafter referred to as S15), of which 25 stars were identified 
as members, and \cite{koposov2015} using VLT/Giraffe with 18 member stars 
(hereafter K15).  Nearly all of the stars in the K15 sample were also included in the 
S15 sample, with the observations were taken 18 days apart.  Hence, only very 
short period binaries with periods less than a year can be expected to show 
significant velocity variations beyond the measurement error between these two datasets.  We also include 
high-resolution velocity measurements of 9 member stars from \cite{ji2016} 
using the Magellan/MIKE spectrograph, along with four additional 
high-resolution measurements from \cite{roederer2016} (hereafter J16/R16).  
Compared to the original S15 dataset, these later measurements occur 226 and 
268 days later, respectively, so these stars may be expected to show velocity 
variations if they are binaries with periods of order a few years or shorter.  
Although we have only eleven stars with repeat measurements over timescales of 
several months, these latter measurements carry greater relative weight due to 
the small measurement errors ($\sim$ 0.1-0.2 km/s) in the J16/R16 datasets.

Since the stars in our sample have velocity measurements from multiple datasets 
using different spectrographs, any systematic offset between the datasets would 
bias the inferred binary fraction in our analysis. To guard against this, we 
will model such systematic offsets explicitly as free parameters. To get a 
sense of whether such offset(s) are present, in Figure \ref{fig:s15_vs_k15} we 
plot a comparison of velocity measurements between the S15 and K15 datasets for 
stars that are common to both. Interestingly, most stars lie well above the 
line $v_{S15} = v_{K15}$, and in a few stars this difference is beyond 
$2\sigma$ for the measurement errors in either dataset. Since the measurements 
were taken only 18 days apart, this apparent offset in a large number of stars 
is unlikely to be explained by binary motion unless the binary population is 
quite extreme.  The implication is that either the K15 data may have a positive 
systematic offset, S15 may have a negative offset, or perhaps some combination 
of the two. Likewise, in Figure \ref{fig:j16_vs_others} we compare the J16 
measurements with S15 and K15. Since the measurements in this figure were taken 
7-8 months apart, we may expect up to a few of the velocity differences to be 
due to binary motion. With this caveat in mind, a few stars in both S15 and K15 
seem to have a positive velocity difference relative to J15 that lies beyond 
2$\sigma$.

With this in mind, we will include offsets in S15 and K15 as free parameters.  
There is little to be gained by including an extra systematic offset parameter 
for the J16 and R16 measurements, because the systemic velocity of the galaxy 
is not known a priori; hence, an offset to all the measured velocities would 
simply translate the resulting mean velocity and thus there would be no unique 
solution for all the offsets and systemic velocity. If it turns out that J16 
and R16 in fact have a systematic offset relative to S16 and K16, our analysis 
would result in a biased inferred systemic velocity, but otherwise this would 
not affect our inferred dispersion or binary fraction.

\begin{figure*}
	\centering
	\includegraphics[height=0.7\hsize]{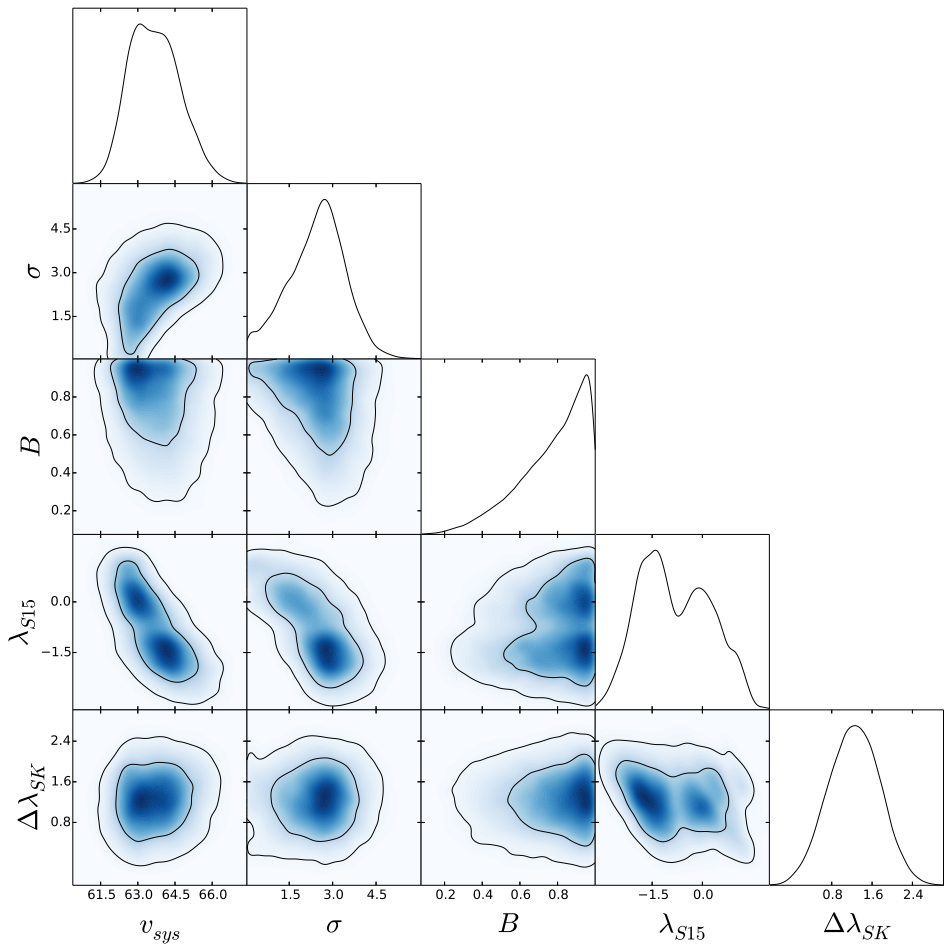}
	\caption{Joint posteriors in the systemic velocity $v_{sys}$, intrinsic 
velocity dispersion $\sigma$, binary fraction $B$, the velocity offset in the
K15 dataset $\lambda_{K15}$, and the relative offset $\Delta\lambda_{SK} = \lambda_{K15}-\lambda_{S15}$. In this analysis, the mean period of Ret II 
is assumed to be $10^4$ days ($\approx$ 27 years), with a spread in the log-normal period distribution identical to Milky Way field binaries $\sigma_{\log P}=2.3$. The posteriors are plotted using a Jeffrey's prior in the dispersion, which is equivalent to having uniform prior in $\ln\sigma$, with uniform priors in the other parameters.}
\label{fig:posts_offset}
\end{figure*}

\section{Results}\label{sec:results}

For our fiducial binary population model, we assume a mean period of $10^4$ days, or roughly 27 years. This is motivated by the work of \cite{Moe2018arXiv180802116M}, who infer this mean period for metal-poor Milky Way halo stars using data from \cite{Latham2002AJ....124.1144L} (see also \cite{Carney2005AJ....129.1886C}). We will then compare our fiducial result to the case of a 180-year mean period, which was inferred for G-dwarf binaries in the solar neighborhood by \cite{duquennoy1991}. For simplicity, in either case we assume a dispersion of log-periods similar to that of Milky Way field binaries ($\sigma_{\log P} = 2.3$). Our velocity offset parameters are defined as the offset in the K15 dataset $\lambda_{K15}$, and the \emph{relative} offset between the S15 and K15 datasets, $\Delta\lambda_{SK} = \lambda_{K15}-\lambda_{S15}$. In our fiducial analysis we assume a Jeffrey's (noninformative) prior in the velocity dispersion, which is equivalent to having a uniform prior in $\ln\sigma$; for all other parameters we assume a uniform prior (which is in fact equivalent to the Jeffrey's prior for those parameters). 

In Figure \ref{fig:posts_offset} we show the resulting posteriors for our fiducial model with mean period of 27 years. Although a large range of binary fractions is allowed 
($B > 0.2$), a high binary fraction is still preferred even for this relatively short mean 
period. The most probable velocity dispersion is 2.8$_{-1.2}^{+0.7}$ km/s, where the uncertainties here denote the 68\% credible interval. There is a small tail in the posterior extending to $\sigma=0$ km/s; note that this tail requires a 
binary fraction $B > 0.8$, since binaries must make up all of the dispersion in this case. In addition, while there is a large uncertainty in 
either of the offsets $\lambda_{S15}$ and $\lambda_{K15}$, we find the relative 
offset $\Delta\lambda_{SK}$ is well constrained to be $1.2 \pm 0.5$ km/s. The 
solution where both offsets are zero lies just outside the 95\% probability 
contours. Hence, there seems to be a clear offset between the K15 and S15 
datasets, but whether one or both of these datasets are offset with respect to the J16 
measurements is quite uncertain given the small number of stars in the J16 
sample.

The metallicity spread of Ret II is already strong evidence that Ret II is a galaxy rather than a globular cluster \citep{willman2012}. Here we check whether its velocity dispersion paints a consistent picture of Ret II being a dark-matter dominated galaxy. If we assume the extreme scenario of \emph{no} dark matter, we can estimate what the expected intrinsic dispersion of Ret II would be. Globular clusters of the Milky Way typically have mass-to-light ratios of 1-2 $M_\odot/L_\odot$. To be conservative, we choose the upper end of this range (2 $M_\odot/L_\odot$) and using Formula 2 in \cite{Wolf2010MNRAS.406.1220W} we find this corresponds to $\sigma \approx 0.21$ km/s. Thus, we define the ``no dark matter'' scenario to be the regime $\sigma \leq 0.21$ km/s, where nearly all of Ret II's dispersion would be due to binaries. For our fiducial 27-year mean period and a noninformative (log) prior in the dispersion, we thus infer a 1\% probability for $M_\odot/L_\odot < 2$; assuming a 180-year mean period (as in solar neighborhood binaries), the probability drops to 0.4\%. In this regime where Ret II's dispersion is due to binaries, the posteriors show there \emph{must} be a positive offset of 1-3 km/s between the K15 velocities and the J16 velocities; in this case, the velocity 
changes due to binaries would be more dramatic, and hence requires a very high 
binary fraction. On the other hand, if the offsets are smaller (K15 $\lesssim$ 1 km/s, 
S15 $\lesssim$ 0 km/s) this does not rule out a high binary fraction, but it does rule 
out the zero-dispersion scenario.

\begin{figure}[t]
	\includegraphics[height=0.8\hsize]{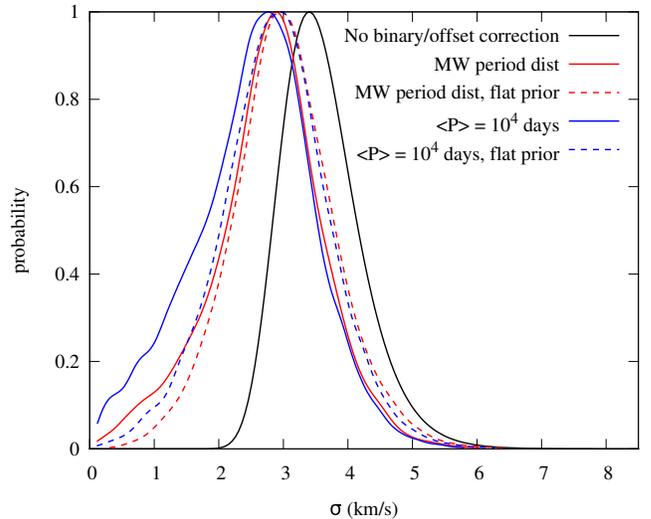}
	\caption{Posteriors in the intrinsic velocity dispersion, assuming different models and priors. The red and blue solid lines include binaries and offset parameters, assuming a log-prior in the velocity dispersion, whereas the corresponding dashed lines assume a uniform prior in the dispersion. Note that the most probable inferred dispersion is fairly robust, peaking around 2.8-2.9 km/s regardless of the binary population and dispersion priors chosen; this is roughly 0.7 km/s lower than the uncorrected estimate (black solid line).}
\label{fig:sigposts}
\end{figure}

How sensitive are our results to the assumed binary population and priors? In Figure \ref{fig:sigposts} we plot the inferred dispersion posterior using a few different models. First, we repeat the analysis but without including binaries or offsets (black line). In this case the most probable inferred dispersion is 3.5 km/s, and there is essentially zero probability of having an intrinsic dispersion less than 2 km/s. We plot our fiducial model with a mean period of 27 years (blue solid line), which used a log-prior in the dispersion; the same model with a uniform prior in the dispersion (blue dotted line) significantly reduces the probability of having a dispersion dominated by binaries, with essentially zero probability of $\sigma\approx 0$ km/s, while the most probable value is relatively unchanged. Finally, we try a model with a mean period of 180 years, similar to G-dwarfs in the Milky Way field, with a log-prior in dispersion (red solid line) as well as a uniform prior (red dashed line). Note that the most probable intrinsic dispersion is fairly robust to these changes in model or priors, in all cases being roughly 0.7 km/s lower compared to the uncorrected dispersion estimate. In addition, the tail going to very small intrinsic dispersions is only significant in our fiducial model (27-year mean period, log-prior in dispersion). With that said, it is certainly possible to increase the odds of having a binary-dominated dispersion if one assumes a shorter mean period; although it seems unlikely, this possibility cannot be ruled out entirely.

\begin{figure*}[ht]
	\centering
	\subfigure[S15 and K15 only]
	{
		\includegraphics[height=0.4\hsize,width=0.48\hsize]{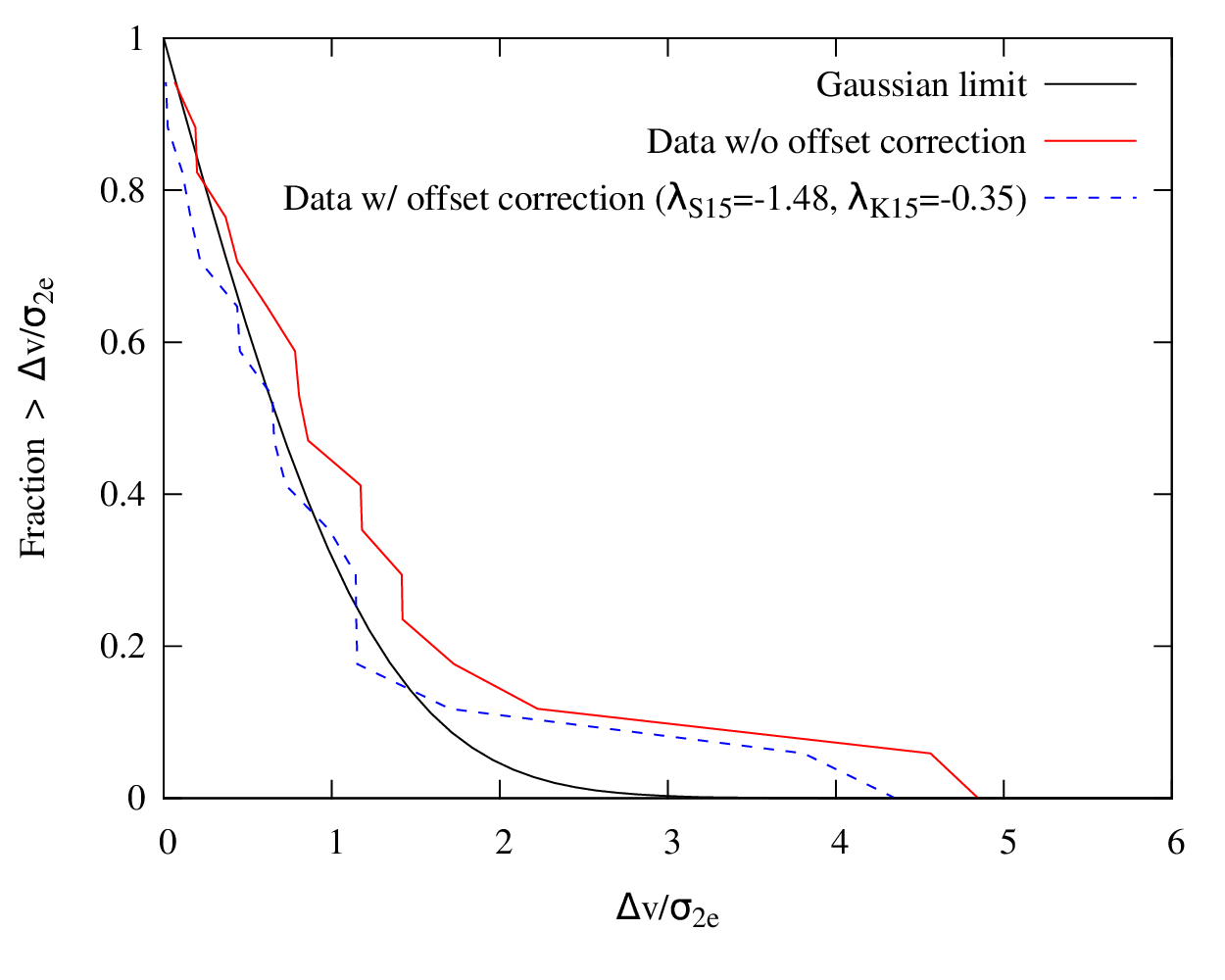}
		\label{fig:dverrchist_short}
	}
	\subfigure[all datasets]
	{
		\includegraphics[height=0.4\hsize,width=0.48\hsize]{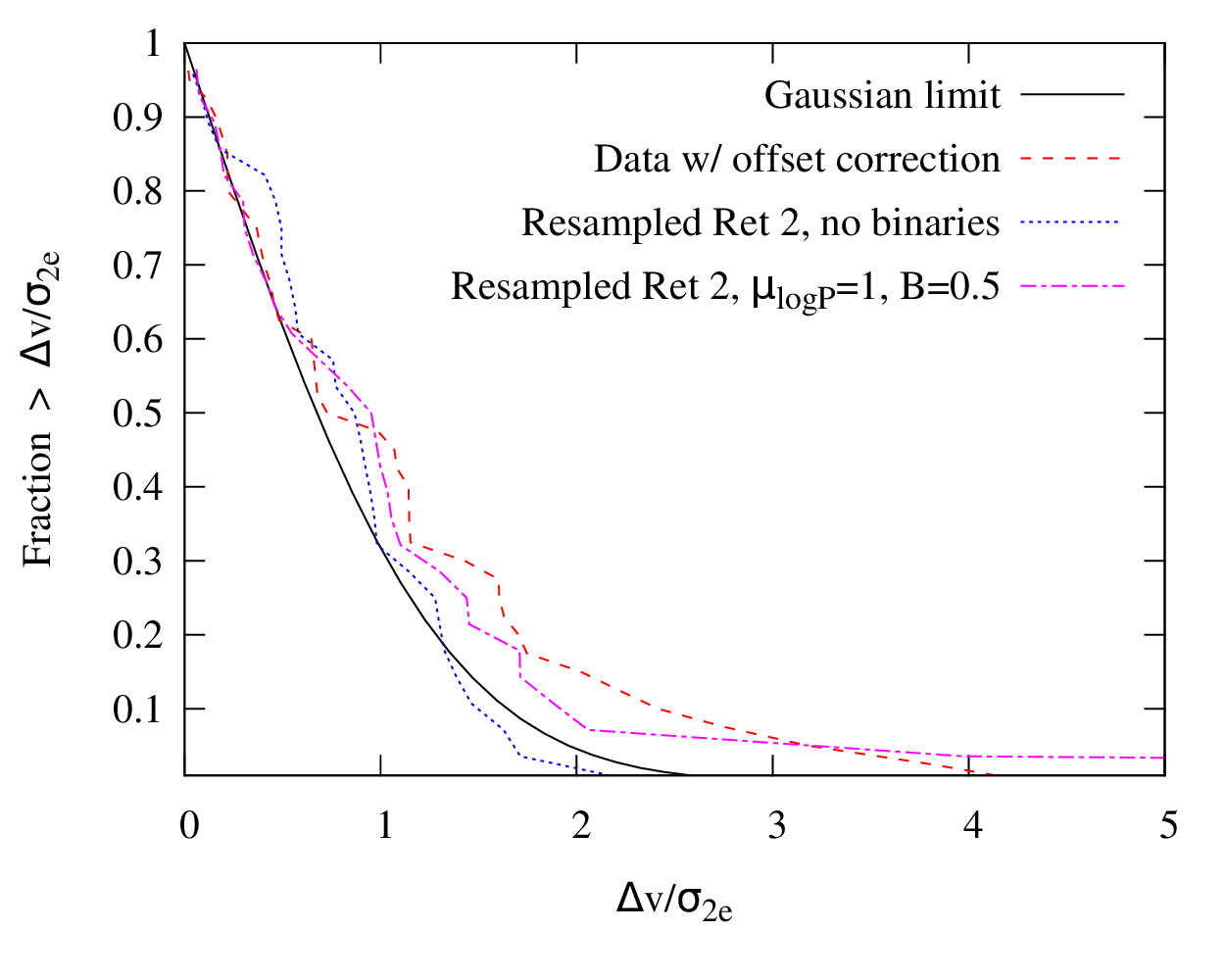}
		\label{fig:dverrchist}
	}
	\caption{The fraction of stars with velocity changes greater than the 
combined measurement error $\sigma_{2e}$, where $\sigma_{2e}^2 = \sigma_1^2 + 
\sigma_2^2$ and $\sigma_1$ and $\sigma_2$ are the measurement errors in the 
first and second measurements respectively. In (a) we only include the S15 and 
K15 measurements, which are separated by only 18 days.  Note that without 
correcting for offsets between datasets, the distribution (red solid curve) 
deviates from the expectation from Gaussian measurement error (solid black 
curve). However, when we apply a correction using our best-fit offset values 
(dashed blue), the distribution more closely follows the Gaussian limit, albeit 
with two outlier stars that are likely to be short-period binaries. In (b) we 
include all the measurements, again showing the data distribution (red dashed 
curve), along with the distribution of randomly resampled velocities assuming 
no binaries (dotted blue curve), and likewise assuming a mean period of 27 years 
and binary fraction of 0.5 (magenta curve).}
\end{figure*}

To check that the inferred offsets are consistent with a Gaussian velocity 
distribution plus binaries, we take the best-fit values for the offsets 
($\lambda_{S15}=-1.48$, $\lambda_{K15}=-0.35$) and apply the correction to the 
the appropriate measurements in the data by subtracting these offsets. In 
Figure \ref{fig:dverrchist_short} we plot the cumulative distribution of 
$\Delta_v/\sigma_{2e}$, where $\Delta_v = |v_1 - v_2|$ is the velocity 
difference between two consecutive epochs for each star and $\sigma_{2e} = 
\sqrt{e_1^2+e_2^2}$ is the corresponding uncertainty in this difference. For 
simplicity, we have only included the S15 and K15 datasets in this figure, for 
which the measurements are only 18 days apart, to minimize the effects of 
binaries. When the offset correction is \emph{not} applied (red solid line), 
the cumulative distribution is systematically above what would be expected for 
a purely Gaussian distribution if the measurement errors properly reflect 
reality (although this is not statistically significant for any individual star 
except for the two that lie beyond 3$\sigma_{2e}$; the latter two have a high 
probability of being short-period binaries). However, when the measurements are 
adjusted for the best-fit offset values, the distribution follows a Gaussian 
quite closely (blue dashed line) with the exception of the two probable binary 
stars. Thus, the best-fit model does indeed appear consistent with a Gaussian 
velocity distribution plus a binary tail due to short period binaries.  We 
should note that the effect of the offsets is not fully captured by the 
deviation of the red curve from Gaussianity here, because it does not indicate 
the sign of $v_1 - v_2$; a random unaccounted-for measurement error could 
produce the same deviation.  However, Figure \ref{fig:s15_vs_k15} clearly 
indicates that most of the S15 measurements are systematically higher than the 
K15 measurements, so we can say with confidence that the correction applied 
here is reasonable.

Next, we plot the distribution of $\Delta_v/\sigma_{2e}$ with all the 
measurements included in Figure \ref{fig:dverrchist}. The red dashed curve 
corresponds to the data with the best-fit offset corrections applied; note 
that, compared to the short-timescale subset we plotted in Figure 
\ref{fig:dverrchist_short}, there are significantly more measurements with 
velocity variations beyond the Gaussian expectation.  Since these extra 
velocity changes occur over longer time scales ($\sim$ 1 year), they can be 
interpreted as being due to binary motion.  This indicates that the two stars 
with large variations over 18 days between the S15 vs. K15 are not the sole 
driver behind the large binary fraction we have inferred. We verified this by 
repeating our analysis with those two stars removed from the sample, and 
found that a high binary fraction is nevertheless preferred by the data as a 
result of the significant velocity changes measured by the J16/R16 data.
Finally, to verify that a galaxy with our best-fit parameters can reasonably 
reproduce the distribution of velocity variations we see in Ret II, we do 
random resamplings of the data and plot the resulting distributions. The blue 
dotted line shows a typical case with a binary fraction of zero; note that the 
large velocity changes in the tail are not well-reproduced, which is typically 
the case for many different random realizations. We then plot a typical case 
with a mean period of 27 years and a binary fraction of 0.5 (meaning half the 
stars in the sample are randomly assigned as binaries, and their binary 
properties are randomly sampled).  It should be emphasized that due to the 
small sample size, the effect of binaries can vary considerably from 
realization to realization, depending on how many close binaries are present; 
in a few cases, there is no discernible binary tail at all. More typically, 
however, a tail exists and in many cases, as shown here, there may be one 
or two stars with velocity changes exceeding 5$\sigma_{2e}$, in some cases as 
large as 30-40 km/s. Such short-period binaries would be likely to have a 
velocity far from the galaxy's systemic velocity at any given time, and hence 
would probably have been flagged as non-member stars. With that caveat aside, 
many realizations do show distributions that are broadly consistent with the 
data.

\begin{figure*}[ht]
	\centering
	\subfigure[$\sigma=0$ km/s]
	{
		\includegraphics[height=0.38\hsize,width=0.48\hsize]{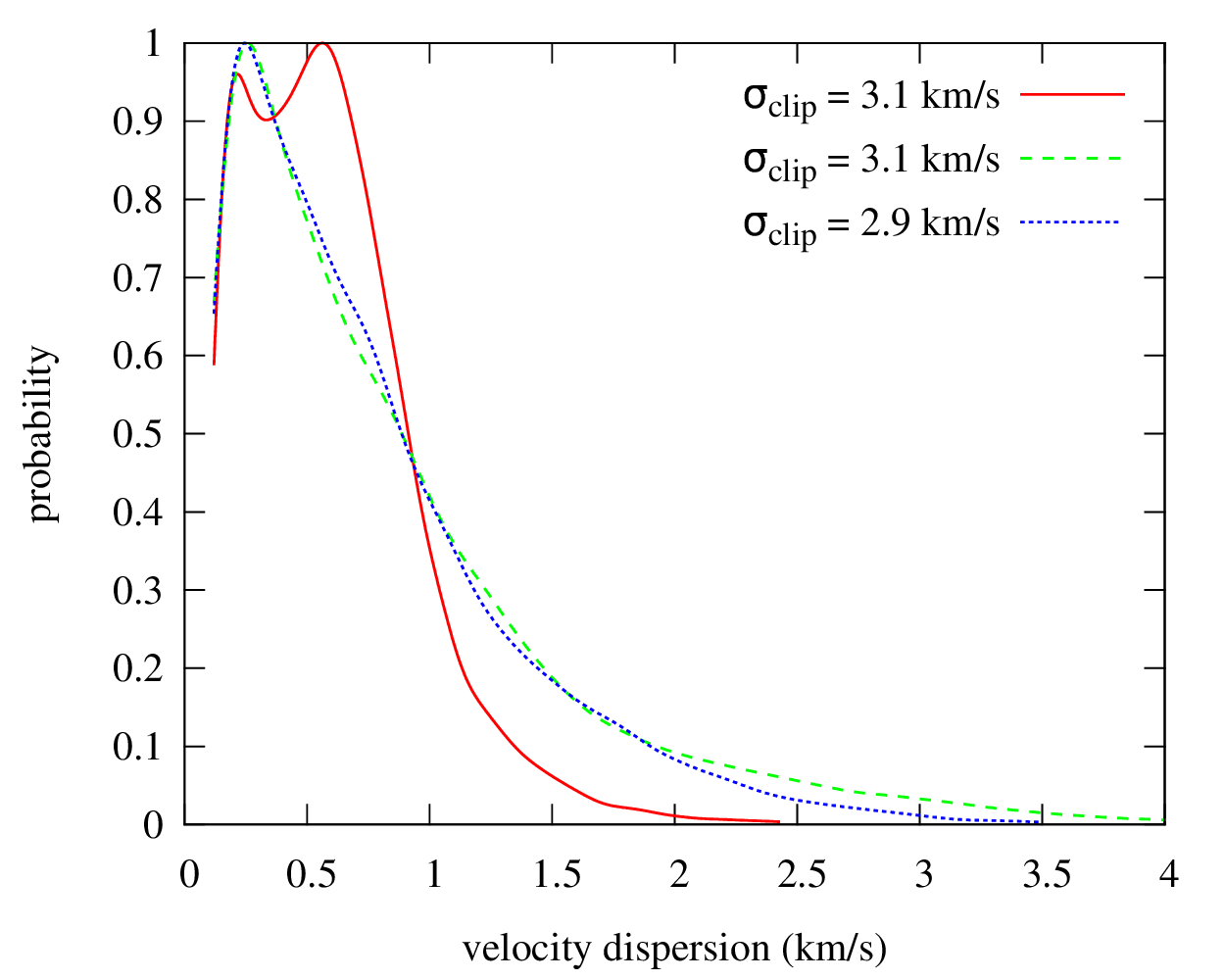}
		\label{fig:sigposts_fake_d0}
	}
	\subfigure[$\sigma=2.7$ km/s]
	{
		\includegraphics[height=0.38\hsize,width=0.48\hsize]{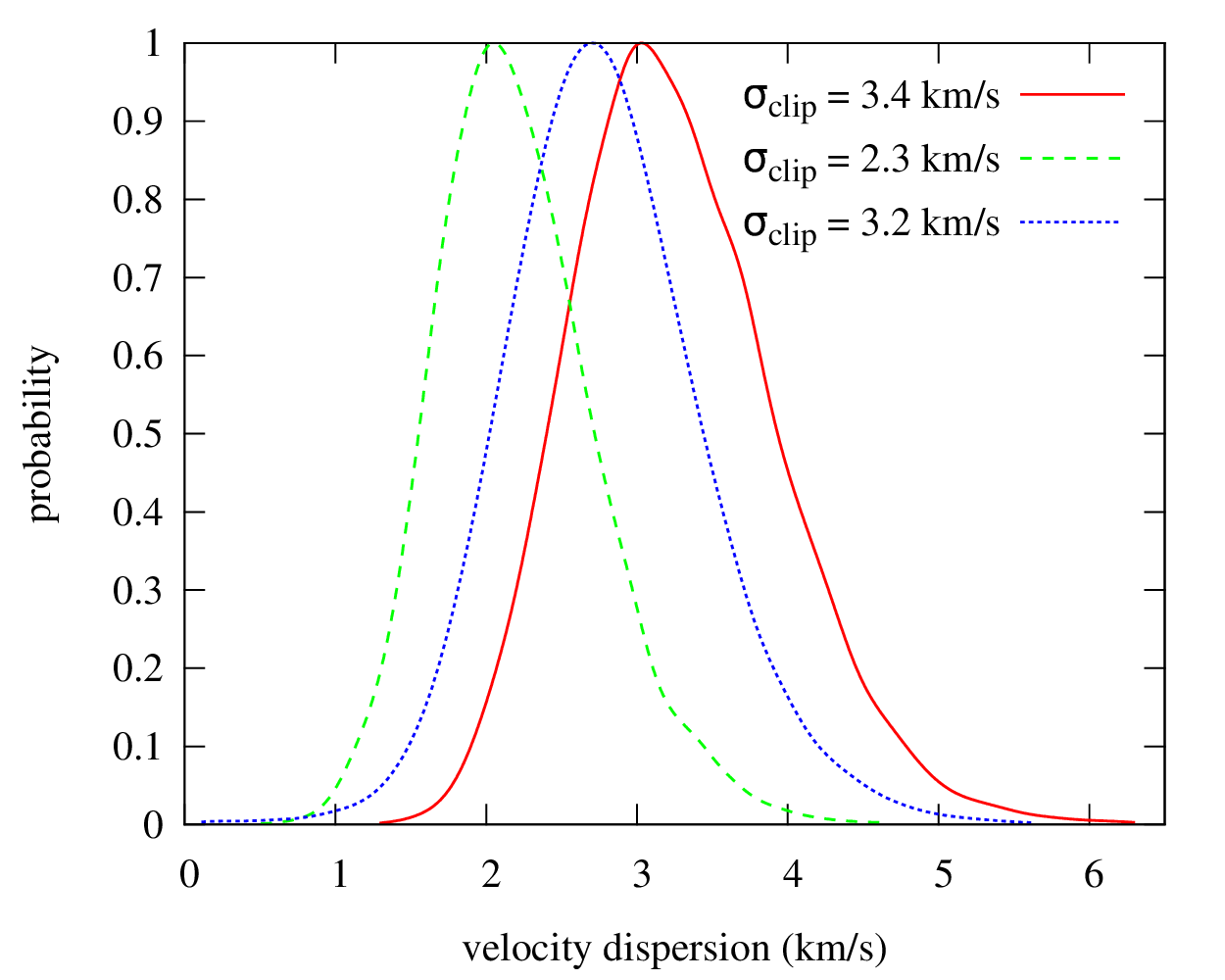}
		\label{fig:sigposts_fake_d2.7}
	}
	\caption{Posteriors in velocity dispersion for simulated datasets where the 
velocity measurements are randomly resampled for each star in Ret II.  In (a) 
we simulate a ``galaxy'' with intrinsic dispersion $\sigma$=0 km/s, while in 
(b) the simulated galaxy has intrinsic dispersion $\sigma$=2.7 km/s. Although 
each case has a high \emph{uncorrected} dispersion $\sigma_{clip}$ determined 
from sigma-clipping, our method can clearly differentiate between these two 
scenarios despite the relatively small sample sizes.}
\end{figure*}

\section{Test of our method on simulated Ret II datasets}

Our analysis shows that although Ret II appears to have a high binary 
fraction, its apparent velocity dispersion of $\sim$ 3 km/s is unlikely to be 
dominated by binary motion unless the measurements of K15 are systematically 
offset from those of J15 by $\approx$ 1-3 km/s. The latter possibility is 
disfavored by the data, but not beyond 95\% confidence limits. This begs the 
question: if the close binary fraction does tend to be high in ultra-faint 
dSphs, given such small samples, can we be confident our method will 
distinguish between galaxies whose dispersion is strongly inflated by binary 
motion versus those that are not?

To investigate this question, we generate mock data by assuming a high binary 
fraction $B=0.9$ and short mean period of 27 years and 
resample all the velocities. We investigate two scenarios: in scenario 1, the 
intrinsic dispersion is $\sigma=0$ km/s (so any apparent velocity dispersion 
would be due entirely to binaries and measurement error); in scenario 2, we 
assume an intrinsic dispersion of $\sigma=2.7$ km/s, which is close to our 
best-fit dispersion of Ret II for the mean period assumed here. For each 
realization, we first calculate the \emph{uncorrected} dispersion using 
3$\sigma$-clipping: we find the maximum likelihood values of $\mu$, $\sigma$, 
then remove stars that lie beyond 3$\sigma$, and repeat as needed until no 
such outliers are present. In order to find cases that roughly resemble Ret II, 
we repeat many random realizations and keep three cases for which the 
uncorrected dispersion $\sigma_{clip}$ lies between 2.5-3.5 km/s. This required 
several realizations before three such cases were found for either 
scenario. In scenario 2 ($\sigma=2.7$ km/s), we did choose one realization with 
$\sigma_{clip} = 2.3$ km/s to see if a nonzero 
intrinsic dispersion can be recovered even for a lower $\sigma_{clip}$ value. After applying our analysis to these mock data, 
the resulting posteriors in the intrinsic dispersion are plotted in Figure 
\ref{fig:sigposts_fake_d0} for the scenario where $\sigma=0$ km/s. In all three 
cases we infer that $\sigma < 1.5$ km/s at greater than 90\% confidence; the 
most probable intrinsic dispersion is well below 1 km/s in all cases. Thus, our 
method appears unlikely to find a peak at $\sigma > 2$ km/s if the apparent velocity 
dispersion is dominated by binaries. In \ref{fig:sigposts_fake_d2.7} we plot 
the resulting posteriors for the cases where the true $\sigma=2.7$ km/s. Here 
we see that $\sigma=0$ km/s is ruled out in all cases; in none of these cases 
do we find a most probable intrinsic dispersion lower than 2 km/s. This builds 
confidence that, despite the relatively small sample size, our binary-corrected 
velocity dispersions can be trusted.

As we mentioned above, despite Ret II having a most probable intrinsic 
dispersion of 3 km/s in our analysis, there is still a nonzero probability of 
$\sigma=0$ km/s. What are we to make of this? First keep in mind that if the 
K15 are measurements are \emph{not} systematically offset from the 
higher-resolution J15 measurements by  $\approx$ 1-3 km/s, the probability 
of Ret II having $\sigma=0$ km/s essentially vanishes. If we retain the 
possibility of a large offset between the K15 and J15 datasets, then there are 
three possible explanations: either (i) Ret II's true intrinsic dispersion is 
in fact somewhere between 0-3 km/s, and the peak appearing at 3 km/s is merely 
due to the small sample size; (ii) Ret II has a close binary fraction 
\emph{even higher} than what we have assumed in the above mock data, and hence 
show velocity variations beyond what the realizations above could produce; or 
(iii) there is some additional systematic in the velocity measurements that is 
producing velocity changes that appear consistent with short-period binary 
motion. Additional spectroscopic measurements will be necessary, to distinguish 
between these scenarios with a high degree of confidence.  Nevertheless, our 
mock data runs bolster confidence that Ret II's apparent dispersion is unlikely 
to be entirely dominated by binary orbital motion.

\section{Discussion}

The result that Ret II's dispersion is unlikely to be entirely due to binaries 
is not surprising, given that its very low metallicities and metallicity spread 
clearly identify it as a dwarf galaxy rather than a globular cluster. However, 
the clear preference for a high binary fraction and/or short mean period in Ret 
II is consistent with a similar result found for the Segue 1 dSph \citep{Martinez2011ApJ...738...55M}. If this 
pattern of a high close binary fraction holds up in other ultra-faints, it 
would be strong evidence that the close binary fraction of a star population is 
a strong function of metallicity, with low metallicity populations harboring a 
greater fraction of close binaries. This anticorrelation between close binary fraction and metallicity has already been pointed out in Milky Way field binaries \citep{Moe2018arXiv180802116M,elbadry2018c,badenes2018}. Whether this is due to a high binary 
fraction, a short mean period, or some combination of the two remains to be 
determined; as discussed in \citet{Minor2013ApJ...779..116M}, the degeneracy between binary 
fraction and the period distribution cannot be broken purely by radial velocity 
measurements without a very large sample with measurements at several epochs. Unfortunately such a large sample is unattainable at present for individual ultra-faint dwarfs.

The most promising approach for breaking the degeneracy between binary fraction and period distribution would be to combine radial velocity measurements with CMD fitting; the latter approach is demonstrated in \cite{Geha2013ApJ...771...29G} and is sensitive only to the binary fraction and stellar masses, not to the periods. An additional independent handle on the binary fraction is possible if main sequence stars are included in the sample, for which binary spectral fitting is possible \citep{elbadry2018,elbadry2018b}. As we have hinted in Section \ref{sec:method}, our method for generating the binary and non-binary likelihoods can be recast in the form of a hierarchical Bayesian model: for individual stars, their orbital 
parameters (e.g. $v_{cm}, P, q$) are marginalized over, while the \emph{prior} 
distributions in these parameters are constrained for the whole population at 
the next level of inference. This approach, which has been used elsewhere to constrain the distribution of orbital parameters in binaries and exoplanets \citep{price2018,foreman2014,hogg2010}, can be incorporated naturally into the methodology used here; in principle, it allows for color-magnitude information and radial velocities to be fit under the same hierarchical framework, and could be applied to a combination of dSph datasets over many epochs to find detailed binary constraints. The demonstration of this more generalized approach in the context of dwarf galaxies is left to future work.

The decrease of the velocity dispersion due to the binary correction has implications for the implied dark matter halo of Ret~II.  
The half-light mass ($M_{1/2}$) of a dispersion supported system is well measured at the half-light radius and it is proportional to $\sigma^2$ \citep{Walker2009ApJ...704.1274W, Wolf2010MNRAS.406.1220W}.
The binary corrected half-light mass will decrease by a factor $\sigma^2$.
Assuming $\sigma=3.3$ \citep{simon2015} we find that the half-mass decreases by a factor of 0.72, decreasing to $M_{1/2}=4.0\times10^5 M_{\odot}$ (from $M_{1/2}=5.6\times10^5 M_{\odot}$).
Searches for dark matter annihilation in ultra-faint dwarfs require an accurate determination of the J-Factor, an integral over the dark matter density squared. 
The J-Factor is proportional to $\sigma^4$  \citep{Pace2018arXiv180206811P} and for Ret~II the  J-Factor will decrease by a factor 0.52. 
Using the J-Factor value from \citet{Pace2018arXiv180206811P}, we find the binary corrected J-Factor to be $\log_{10}{\left(J/{\rm GeV^2 \, cm^{-5} }\right)}=18.58$ (from 18.87).

\section{Conclusion}

We have applied a Bayesian method to model multi-epoch radial velocities 
in the ultra-faint dwarf Reticulum II, fully accounting for the 
effects of binary orbital motion and systematic offsets between different 
spectroscopic datasets. Our primary results are encapsulated in Figures \ref{fig:posts_offset} and \ref{fig:sigposts}, where we infer the intrinsic dispersion and binary fraction of Ret II. Despite the relatively small sample size (26 stars in 
total), we find that the binary fraction of Ret II is higher than 0.5 at the 
90\% confidence level, if the mean orbital period is assumed to be 30 years or 
longer. Despite this high binary fraction, the best-fit intrinsic dispersion of 
Ret II is 2.8$_{-1.2}^{+0.7}$ km/s, consistent with Ret II having a large 
mass-to-light ratio. Our best-fit velocity dispersion is smaller than previous estimates and implies a weaker dark matter annihilation signal, with the J-factor reduced by a factor of $\approx$ 0.5 compared to the results of \cite{simon2015}. Assuming a mean period of $10^4$ days (which was recently inferred in low-metallicity Milky Way binaries in \citealt{Moe2018arXiv180802116M}), we estimate a $\approx$ 1\% probability that Ret II's dispersion is due to binaries rather than dark matter, corresponding to the regime $M_{\odot}/{L_\odot} \approx$ 2. These results are thus consistent with Ret II being a dark matter-dominated galaxy to high significance, in agreement with the expectation from its large metallicity dispersion.

Beyond the importance of obtaining unbiased mass estimates of ultra-faint 
dwarfs, binary populations in these objects are interesting in their own right 
as they may hold clues to star formation in extremely metal-poor environments.  
The fact that Ret II appears to have a high close binary fraction is consistent with 
previous results from the Segue 1 ultra-faint dwarf, and echoes similar results 
from Milky Way halo stars that suggest that metal-poor star populations have a 
relatively high fraction of close binaries. A more robust and detailed 
characterization of binary populations in dwarf galaxies will require a larger 
multi-epoch sample for a large number of dwarfs, a combination of deep 
photometry and high-resolution spectroscopy, and the application of a fully 
hierarchical version of the Bayesian method we have applied in this paper. Over 
the long term, the binary populations in these extreme objects might hold vital 
clues to a deeper understanding of the physics underlying star formation.

\section*{Acknowledgements}

We gratefully acknowledge a grant of computer time from XSEDE allocation 
TG-AST130007.
This research was also supported, in part, by a grant of computer time from the City 
University of New York High Performance Computing Center under NSF Grants 
CNS-0855217, CNS-0958379 and ACI-1126113.
QM was supported by NSF grant AST-1615306. L.E.S acknowledges support from DOE Grant de-sc0010813. A.B.P., J.L.M., and L.E.S. acknowledge generous support from the George P. and Cynthia Woods Institute for Fundamental Physics and Astronomy at Texas A\&M University. 

This work was supported in part by NSF grant AST-1153335.

~ ~ ~ ~ ~ ~ ~ ~ ~ ~ ~ ~ ~ ~ ~ ~ ~ ~ ~ ~ ~ ~ ~ ~ ~ ~ ~ ~ ~ ~ ~ ~ ~ ~ ~ ~ ~ ~ ~

\bibliographystyle{apj}
\bibliography{binary4}

\begin{thebibliography}{49}
\expandafter\ifx\csname natexlab\endcsname\relax\def\natexlab#1{#1}\fi

\bibitem[{{Ackermann} {et~al.}(2015)}]{ackermann2015}
{Ackermann}, M. {et~al.} 2015, Physical Review Letters, 115, 231301

\bibitem[{{Badenes} {et~al.}(2018)}]{badenes2018}
{Badenes}, C. {et~al.} 2018, \apj, 854, 147

\bibitem[{{Bechtol} {et~al.}(2015)}]{Bechtol2015ApJ...807...50B}
{Bechtol}, K. {et~al.} 2015, \apj, 807, 50

\bibitem[{{Brown} {et~al.}(2012)}]{brown2012}
{Brown}, T.~M. {et~al.} 2012, \apjl, 753, L21

\bibitem[{{Carney} {et~al.}(2005){Carney}, {Aguilar}, {Latham}, \&
  {Laird}}]{Carney2005AJ....129.1886C}
{Carney}, B.~W., {Aguilar}, L.~A., {Latham}, D.~W., \& {Laird}, J.~B. 2005,
  \aj, 129, 1886

\bibitem[{{Drlica-Wagner} {et~al.}(2015)}]{Drlica-Wagner2015ApJ...809L...4D}
{Drlica-Wagner}, A. {et~al.} 2015, \apjl, 809, L4

\bibitem[{{Duquennoy} \& {Mayor}(1991)}]{duquennoy1991}
{Duquennoy}, A. \& {Mayor}, M. 1991, \aap, 248, 485

\bibitem[{{El-Badry} \& {Rix}(2018)}]{elbadry2018c}
{El-Badry}, K. \& {Rix}, H.-W. 2018, ArXiv e-prints

\bibitem[{{El-Badry} {et~al.}(2018{\natexlab{a}}){El-Badry}, {Rix}, {Ting},
  {Weisz}, {Bergemann}, {Cargile}, {Conroy}, \& {Eilers}}]{elbadry2018}
{El-Badry}, K., {Rix}, H.-W., {Ting}, Y.-S., {Weisz}, D.~R., {Bergemann}, M.,
  {Cargile}, P., {Conroy}, C., \& {Eilers}, A.-C. 2018{\natexlab{a}}, \mnras,
  473, 5043

\bibitem[{{El-Badry} {et~al.}(2018{\natexlab{b}})}]{elbadry2018b}
{El-Badry}, K. {et~al.} 2018{\natexlab{b}}, \mnras, 476, 528

\bibitem[{{Elbert} {et~al.}(2018){Elbert}, {Bullock}, {Kaplinghat},
  {Garrison-Kimmel}, {Graus}, \& {Rocha}}]{elbert2018}
{Elbert}, O.~D., {Bullock}, J.~S., {Kaplinghat}, M., {Garrison-Kimmel}, S.,
  {Graus}, A.~S., \& {Rocha}, M. 2018, \apj, 853, 109

\bibitem[{{Foreman-Mackey} {et~al.}(2014){Foreman-Mackey}, {Hogg}, \&
  {Morton}}]{foreman2014}
{Foreman-Mackey}, D., {Hogg}, D.~W., \& {Morton}, T.~D. 2014, \apj, 795, 64

\bibitem[{{Geha} {et~al.}(2013)}]{Geha2013ApJ...771...29G}
{Geha}, M. {et~al.} 2013, \apj, 771, 29

\bibitem[{{Hogg} {et~al.}(2010){Hogg}, {Myers}, \& {Bovy}}]{hogg2010}
{Hogg}, D.~W., {Myers}, A.~D., \& {Bovy}, J. 2010, \apj, 725, 2166

\bibitem[{{Ji} {et~al.}(2016{\natexlab{a}}){Ji}, {Frebel}, {Simon}, \&
  {Chiti}}]{ji2016}
{Ji}, A.~P., {Frebel}, A., {Simon}, J.~D., \& {Chiti}, A. 2016{\natexlab{a}},
  \apj, 830, 93

\bibitem[{{Ji} {et~al.}(2016{\natexlab{b}}){Ji}, {Frebel}, {Simon}, \&
  {Geha}}]{Ji2016ApJ...817...41J}
{Ji}, A.~P., {Frebel}, A., {Simon}, J.~D., \& {Geha}, M. 2016{\natexlab{b}},
  \apj, 817, 41

\bibitem[{{Kim} \& {Jerjen}(2015)}]{Kim2015ApJ...808L..39K}
{Kim}, D. \& {Jerjen}, H. 2015, \apjl, 808, L39

\bibitem[{{Kirby} {et~al.}(2015){Kirby}, {Cohen}, {Simon}, \&
  {Guhathakurta}}]{Kirby2015ApJ...814L...7K}
{Kirby}, E.~N., {Cohen}, J.~G., {Simon}, J.~D., \& {Guhathakurta}, P. 2015,
  \apjl, 814, L7

\bibitem[{{Kirby} {et~al.}(2017){Kirby}, {Cohen}, {Simon}, {Guhathakurta},
  {Thygesen}, \& {Duggan}}]{Kirby2017ApJ...838...83K}
{Kirby}, E.~N., {Cohen}, J.~G., {Simon}, J.~D., {Guhathakurta}, P., {Thygesen},
  A.~O., \& {Duggan}, G.~E. 2017, \apj, 838, 83

\bibitem[{{Kirby} {et~al.}(2008){Kirby}, {Simon}, {Geha}, {Guhathakurta}, \&
  {Frebel}}]{kirby2008}
{Kirby}, E.~N., {Simon}, J.~D., {Geha}, M., {Guhathakurta}, P., \& {Frebel}, A.
  2008, \apjl, 685, L43

\bibitem[{{Koch} {et~al.}(2009){Koch}, {Wilkinson}, {Kleyna}, {Irwin},
  {Zucker}, {Belokurov}, {Gilmore}, {Fellhauer}, \&
  {Evans}}]{Koch2009ApJ...690..453K}
{Koch}, A., {Wilkinson}, M.~I., {Kleyna}, J.~T., {Irwin}, M., {Zucker}, D.~B.,
  {Belokurov}, V., {Gilmore}, G.~F., {Fellhauer}, M., \& {Evans}, N.~W. 2009,
  \apj, 690, 453

\bibitem[{{Koposov} {et~al.}(2015{\natexlab{a}}){Koposov}, {Belokurov},
  {Torrealba}, \& {Evans}}]{Koposov2015ApJ...805..130K}
{Koposov}, S.~E., {Belokurov}, V., {Torrealba}, G., \& {Evans}, N.~W.
  2015{\natexlab{a}}, \apj, 805, 130

\bibitem[{{Koposov} {et~al.}(2015{\natexlab{b}})}]{koposov2015}
{Koposov}, S.~E. {et~al.} 2015{\natexlab{b}}, \apj, 811, 62

\bibitem[{{Latham} {et~al.}(2002){Latham}, {Stefanik}, {Torres}, {Davis},
  {Mazeh}, {Carney}, {Laird}, \& {Morse}}]{Latham2002AJ....124.1144L}
{Latham}, D.~W., {Stefanik}, R.~P., {Torres}, G., {Davis}, R.~J., {Mazeh}, T.,
  {Carney}, B.~W., {Laird}, J.~B., \& {Morse}, J.~A. 2002, \aj, 124, 1144

\bibitem[{{Li} {et~al.}(2018)}]{Li2018ApJ...857..145L}
{Li}, T.~S. {et~al.} 2018, \apj, 857, 145

\bibitem[{{Lovell} {et~al.}(2014){Lovell}, {Frenk}, {Eke}, {Jenkins}, {Gao}, \&
  {Theuns}}]{lovell2014}
{Lovell}, M.~R., {Frenk}, C.~S., {Eke}, V.~R., {Jenkins}, A., {Gao}, L., \&
  {Theuns}, T. 2014, \mnras, 439, 300

\bibitem[{{Luque} {et~al.}(2016)}]{Luque2016MNRAS.458..603L}
{Luque}, E. {et~al.} 2016, \mnras, 458, 603

\bibitem[{{Luque} {et~al.}(2017{\natexlab{a}})}]{Luque2017arXiv170905689L}
---. 2017{\natexlab{a}}, ArXiv e-prints

\bibitem[{{Luque} {et~al.}(2017{\natexlab{b}})}]{Luque2017MNRAS.468...97L}
---. 2017{\natexlab{b}}, \mnras, 468, 97

\bibitem[{{Martin} {et~al.}(2016)}]{Martin2016ApJ...818...40M}
{Martin}, N.~F. {et~al.} 2016, \apj, 818, 40

\bibitem[{{Martinez} {et~al.}(2011){Martinez}, {Minor}, {Bullock},
  {Kaplinghat}, {Simon}, \& {Geha}}]{Martinez2011ApJ...738...55M}
{Martinez}, G.~D., {Minor}, Q.~E., {Bullock}, J., {Kaplinghat}, M., {Simon},
  J.~D., \& {Geha}, M. 2011, \apj, 738, 55

\bibitem[{{McConnachie} \&
  {C{\^o}t{\'e}}(2010)}]{McConnachie2010ApJ...722L.209M}
{McConnachie}, A.~W. \& {C{\^o}t{\'e}}, P. 2010, \apjl, 722, L209

\bibitem[{{Minor}(2013)}]{Minor2013ApJ...779..116M}
{Minor}, Q.~E. 2013, \apj, 779, 116

\bibitem[{{Minor} {et~al.}(2010){Minor}, {Martinez}, {Bullock}, {Kaplinghat},
  \& {Trainor}}]{Minor2010ApJ...721.1142M}
{Minor}, Q.~E., {Martinez}, G., {Bullock}, J., {Kaplinghat}, M., \& {Trainor},
  R. 2010, \apj, 721, 1142

\bibitem[{{Moe} {et~al.}(2018){Moe}, {Kratter}, \&
  {Badenes}}]{Moe2018arXiv180802116M}
{Moe}, M., {Kratter}, K.~M., \& {Badenes}, C. 2018, ArXiv e-prints

\bibitem[{{Olszewski} {et~al.}(1996){Olszewski}, {Pryor}, \&
  {Armandroff}}]{olszewski1996}
{Olszewski}, E.~W., {Pryor}, C., \& {Armandroff}, T.~E. 1996, \aj, 111, 750

\bibitem[{{Pace} \& {Strigari}(2018)}]{Pace2018arXiv180206811P}
{Pace}, A.~B. \& {Strigari}, L.~E. 2018, ArXiv e-prints

\bibitem[{{Price-Whelan} {et~al.}(2018)}]{price2018}
{Price-Whelan}, A.~M. {et~al.} 2018, \aj, 156, 18

\bibitem[{{Raghavan} {et~al.}(2010)}]{raghavan2010}
{Raghavan}, D. {et~al.} 2010, \apjs, 190, 1

\bibitem[{{Rocha} {et~al.}(2013){Rocha}, {Peter}, {Bullock}, {Kaplinghat},
  {Garrison-Kimmel}, {O{\~n}orbe}, \& {Moustakas}}]{rocha2013}
{Rocha}, M., {Peter}, A.~H.~G., {Bullock}, J.~S., {Kaplinghat}, M.,
  {Garrison-Kimmel}, S., {O{\~n}orbe}, J., \& {Moustakas}, L.~A. 2013, \mnras,
  430, 81

\bibitem[{{Roederer} {et~al.}(2016)}]{roederer2016}
{Roederer}, I.~U. {et~al.} 2016, \aj, 151, 82

\bibitem[{{Simon} \& {Geha}(2007)}]{simon2007}
{Simon}, J.~D. \& {Geha}, M. 2007, \apj, 670, 313

\bibitem[{{Simon} {et~al.}(2011)}]{Simon2011ApJ...733...46S}
{Simon}, J.~D. {et~al.} 2011, \apj, 733, 46

\bibitem[{{Simon} {et~al.}(2015)}]{simon2015}
---. 2015, \apj, 808, 95

\bibitem[{{Spencer} {et~al.}(2017){Spencer}, {Mateo}, {Walker}, {Olszewski},
  {McConnachie}, {Kirby}, \& {Koch}}]{spencer2017}
{Spencer}, M.~E., {Mateo}, M., {Walker}, M.~G., {Olszewski}, E.~W.,
  {McConnachie}, A.~W., {Kirby}, E.~N., \& {Koch}, A. 2017, \aj, 153, 254

\bibitem[{{Walker} {et~al.}(2009){Walker}, {Mateo}, {Olszewski},
  {Pe{\~n}arrubia}, {Wyn Evans}, \& {Gilmore}}]{Walker2009ApJ...704.1274W}
{Walker}, M.~G., {Mateo}, M., {Olszewski}, E.~W., {Pe{\~n}arrubia}, J., {Wyn
  Evans}, N., \& {Gilmore}, G. 2009, \apj, 704, 1274

\bibitem[{{Wilkinson} {et~al.}(2002){Wilkinson}, {Kleyna}, {Evans}, \&
  {Gilmore}}]{wilkinson2002}
{Wilkinson}, M.~I., {Kleyna}, J., {Evans}, N.~W., \& {Gilmore}, G. 2002,
  \mnras, 330, 778

\bibitem[{{Willman} \& {Strader}(2012)}]{willman2012}
{Willman}, B. \& {Strader}, J. 2012, \aj, 144, 76

\bibitem[{{Wolf} {et~al.}(2010){Wolf}, {Martinez}, {Bullock}, {Kaplinghat},
  {Geha}, {Mu{\~n}oz}, {Simon}, \& {Avedo}}]{Wolf2010MNRAS.406.1220W}
{Wolf}, J., {Martinez}, G.~D., {Bullock}, J.~S., {Kaplinghat}, M., {Geha}, M.,
  {Mu{\~n}oz}, R.~R., {Simon}, J.~D., \& {Avedo}, F.~F. 2010, \mnras, 406, 1220

\end{thebibliography}


\end{document}